\documentclass[12pt]{article}
\usepackage{booktabs, dcolumn}
\usepackage[T1]{fontenc}
\usepackage[utf8]{inputenc}
\usepackage{geometry}
\usepackage{color}
\usepackage{amsmath}
\usepackage{bm}
\usepackage{setspace}
\usepackage[authoryear]{natbib}
\usepackage[font=footnotesize, width=1\textwidth]{caption}
\usepackage{graphicx}
\usepackage{hyperref}
\usepackage{times}
\usepackage{pgfplots}

\usepgfplotslibrary{dateplot}
\usepgfplotslibrary[statistics]

\pgfplotsset{compat=1.5}

\newcommand{\ra}[1]{\renewcommand{\arraystretch}{#1}}
\newcommand{\mc}[1]{\multicolumn{1}{c@{}}{#1}}
\newcolumntype{d}[1]{D{.}{.}{#1}}

\newcommand{\ie}{\textit{i.e.}}
\newcommand{\exante}{\textit{ex-ante}}
\newcommand{\wrt}{w.r.t.}

\begin{document}

\title{Why and how systematic strategies decay
\thanks{We thank Yves Lemperi\`ere, Philip Seager, Mark Potters and Jean-Philippe Bouchaud for helpful feedback at an early stage of this project.}
\bigskip
} 
\vspace{.5cm}

\author{
  Antoine Falck\thanks{Capital Fund Management Inc, The Chrysler Building, 405 Lexington Avenue -- 55th Fl., New York, NY, 10174}
  \and Adam Rej\footnotemark[2]
  \and David Thesmar\footnotemark[2] \thanks{MIT, NBER \& CEPR}
}

\maketitle

\begin{abstract}

In this paper, we propose \exante{} characteristics that predict the drop in risk-adjusted performance out-of-sample for a large set of stock anomalies published in finance and accounting academic journals.
Our set of predictors is generated by hypotheses of OOS decay put forward by \cite{mclean2016}: arbitrage capital flowing into newly published strategies and in-sample overfitting linked to multiple hypothesis testing.
The year of publication alone---compatible with both hypotheses---explains 30\% of the variance of Sharpe decay across factors: Every year, the Sharpe decay of newly-published factors increases by 5ppt.
The other important variables are directly related to overfitting: the number of operations required to calculate the signal and two measures of sensitivity of in-sample Sharpe to outliers together add another 15\% of explanatory power.
Some arbitrage-related variables are statistically significant, but their predictive power is marginal.  

\end{abstract}

\newpage

\section{Introduction}

This paper investigates the determinants of the performance decay of known investment strategies after their publication. Recently, several authors have conducted a census and a subsequent analysis of the literature documenting stock returns predictability (\cite{harvey2015, mclean2016, hou2018, Chen} among others). The overwhelming finding of the these studies is that results significantly deteriorate outside the in-sample period. Using a sample of 72 published investment strategies, we first replicate this observation and find, in line with existing studies, that the Sharpe ratio of these strategies declines by about one half after the results are published. We also study the performance of these factors on international pools of stocks (as in \cite{asness2013}, \cite{RobStambaugh} and more recently \cite{revisionist}). 

In this paper, we isolate \exante{} characteristics that predict the drop in risk-adjusted performance out-of-sample.  Why would performance decay? \cite{mclean2016} suggest two possibilities: arbitrage or overfitting. Arbitrage arises because, as soon as new investment idea is disseminated, arbitrage capital moves in, reducing trading profits. This is a concern for investors trading on known ideas (published research), but much less so for investors using proprietary ideas that only slowly leak out of their organizations. Of much greater concern is overfitting, \ie{} the notion that a portion of the risk-adjusted performance only happened by chance because the researcher has tested multiple hypotheses (as discussed by \cite{harvey2015} for instance). These two views suggest a wealth of potential drivers of out-of-sample decay. We build \textit{four} proxies of ease of arbitrage, \textit{six} proxies signaling overfitting and we complete the list with the publication date, which may be associated with both mechanisms. Among these variables, some have significant predictive power. The year of publication alone explains 30\% of the variance of Sharpe decay across factors: Every year, the Sharpe decay of \textit{newly-published} factors increases by 5ppt. The other variables are: the number of operations required to calculate the signal and two measures of sensitivity of in-sample Sharpe to outliers together add another 15\% of explanatory power. Arbitrage-related variables only marginally contribute.  

\bigskip

This article is structured as follows. We first describe the set of 72 known stock return anomalies that we have replicated using our own data. Each anomaly is based on information from CRSP or COMPUSTAT (US listed firms).

Using this set of anomalies, we confirm the finding by \cite{mclean2016} that performance of signals drops significantly after the publication date. We calculate the in-sample (IS) Sharpe ratio of a long-short strategy based on each signal using all the data available to the authors, as well Sharpe ratios in several ``validation samples''. First, in the spirit of \cite{mclean2016}, we show that Sharpe ratios of strategies decay by about one half \emph{after} publication---this magnitude is consistent with their finding. Second, we look at other countries (as in \cite{asness2013},\cite{RobStambaugh} and more recently \cite{revisionist}). All of the factors we reconstructed were discovered on US data (typically the CRSP universe minus penny stocks), so a natural question is whether these findings still hold in other countries. International stocks are not a perfect out-of-sample test, as some of the predictability may be US-specific and may therefore not translate into similar predictability in other countries---for instance because of accounting or regulatory differences. Another challenge is that international pools, conditional on liquidity, tend to be considerably smaller than the U.S. stock market. We thus size-adjust Sharpe ratios to compare performance, as large stock pools allow for more diversification. We propose two simple methods to implement a ``fair'' comparison across pools. We find that without such an adjustment the drop in Sharpe looks considerable larger than the within-US post-publication drop: on average about 90\% across pools. With size adjustments in place, we find haircuts in the range 25\% - 50\% -- a magnitude similar to the post-publication decay.

We then construct 11 different variables that are expected to predict post-publication Sharpe decay: four of them related to arbitrage and six pertaining to overfitting. We further add the date of publication, which is related to both views. These proxies are related to one another so we first study their predictive power separately.

Out of the 11 variables we investigate, six variables are found to predict out-of-sample Sharpe decay. The main predictor of decay is the date of publication. More recently published factors perform less well out-of-sample. Quantitatively, we find that every decade the applicable haircut decreases by approximately 50\%.  This is consistent with the idea that recently published signals are more likely to have been mined, but also that arbitrage capital ``rushes in'' faster after publication in recent years. Consistent with the mining hypothesis, we document that the complexity of predictor has increased over time, and that performance decays very fast after publication, but we are aware this is only suggestive evidence.

The next two variables are inspired by the arbitrage view and are related with one another: they measure the relative market capitalization of stocks actively traded by the factor. When the traded firms are large, the performance drop should be more pronounced. Large companies are in general cheaper to trade and allow for larger trading capacity. Thus these two predictors are consistent with the idea that arbitrage capital tends to flow in more aggressively into anomalies that have enough capacity. 

The last three predictors are more directly related to overfitting. First, we find that model flexibility, as proxied by the number of operations involved in computing the predicting variable, predicts OOS performance decay. This is consistent with researchers mining for more complex relationships to improve the in-sample backtest. Next come two overfitting variables that are related to bias arising from small sample bias in the  training data. The first variable is the ``sensitivity to removing a small subset of stocks''. We construct this measure by computing the range of Sharpe ratios obtained in sample by removing randomly 10\% of the pool over 100 draws. When the range is large, the in-sample performance of the factor is vulnerable to a small number of stocks. The second variable is directly inspired by \cite{broderick2020}: a new version of factors which uses all of the in-sample observations, save for the 0.1\% observations that are most impactful. We find that both vulnerability and sensitivity significantly predict the decay of performance after publication. All of these variables significantly predict out of Sharpe decay. 

Last, in order to compare the explanatory powers of the two sets of variables, we aggregate the arbitrage and  overfitting variables into two ``global'' measures. We regress the Sharpe decay on these two aggregates and the date of publication. We find that the date of publication alone predicts 30\% of the variance of decays and the overfitting variables another 15\%. The additional contribution of arbitrage-related variables is negligible.

\bigskip

This paper is mostly a contribution to the rising literature on robustness of findings in economics in general. An earlier literature documented the accumulation of empirical results around key thresholds of statistical significance, as an evidence that many empirical results are fitted to ``work'' in-sample (see \cite{EmpireStrickesBack} for a recent reference). Some papers advocate or develop methods to deal with multiple testing problems (see for instance \cite{harvey2015}, \cite{sterling} or \cite{giglio_square} for recent work). Other papers address the issue of robustness more directly. \cite{revisionist} propose a Bayesian-motivated shrinkage of in-sample alphas towards zero, which attenuates out-of-sample decay. \cite{broderick2020} propose a method to estimate the contribution of the most influential points. Our paper proposes in a way an out-of-sample test of these methodologies: the quest for returns predictability has generated a very large literature of comparable findings. This makes it easier to see if, on a large cross-section of such findings, there are early warning signs of overfitting. Contrary to our expectations, we find that the in-sample $t$-stat is only a weak predictor of performance decay. Indicators of sensitivity to a small number of observations (such as the one inspired by \cite{broderick2020}) are more successful as early indicators of out of sample decay.

\bigskip
\section{Data}

\subsection{Data Sources}

In this paper, we focus on characteristics-based factor returns discovered in the finance literature. Most of this research exploits U.S. data on prices and accounting information. Hence, our main source of data is an extract of the CRSP-COMPUSTAT merged sample that runs from January 1963 to April 2014. We also use CFM's proprietary data for international stocks. These international data have a smaller range. The beginning date depends on the pool, as shown in Table \ref{tab:index_date}. The end date is December 2018 for all pools. International data will serve as out-of-sample tests of robustness for the various factors discovered on U.S. data, an approach that differs from the post-publication period of \cite{mclean2016}.

\begin{table}[htbp]
  \centering
  \ra{1.3}
   \caption{Start dates of  accounting and price data sets for different international indexes.}
   \label{tab:index_date}
  \begin{tabular}{@{}ll@{}}
    \toprule
    \mc{\textbf{Index name}}       & \mc{\textbf{Start date}} \\
   \midrule
    S\&P/ASX 200 Index        & 2001-11-01 \\
    Bloomberg European 500    & 1999-01-01 \\
    Hang Seng Composite Index & 2002-01-01 \\
    Korea Kospi Index         & 2003-12-01 \\
    Russell 3000 Index        & 1995-01-01 \\
    Russell 1000 Index        & 1995-01-01 \\
    China SE Shang Composite  & 2002-01-01 \\
    S\&P/TSX Composite Index  & 2001-04-01 \\
    TOPIX 500 Index (TSE)     & 2007-01-01 \\
    FTSE 100 Index            & 1996-01-01 \\
   \bottomrule
   \end{tabular}
\end{table}

\subsection{Replication Approach}

We construct a set of 72 factors documented in the literature (see list in appendix table \ref{tab:list_factor}) and mark them to market on a daily basis. We restrict our literature review to characteristics based on COMPUSTAT or CRSP only. Factors that were published after 2010 are not added to our ``zoo'', as our dataset stops in 2014 and we want to have a decent out-of-sample period. The size of our ``zoo'' is in the range of existing factor studies (50 factors in \cite{kozak2018} and \cite{valentin}, 97 factors in \cite{mclean2016}, and 452 in \cite{hou2018}). 

\cite{hou2018} distinguish three types of replication: pure (same method, same sample), statistical (same method, different sample) and scientific (similar method, different sample), which is closer to our approach. We use either the same sample (in-sample), or alternative samples (post-publication, international pools, liquid pools of the 500$^{th}$ and 1,000$^{th}$ largest stocks). We use similar, but slightly different, methodologies: We construct predictors by staying as close as possible to the papers, but improve marginally on the portfolio construction. Let us explain. For each predictor, the construction of the portfolio follows the publication as closely as possible. We use the same variable, the same formulas as well as the same ranking procedure---most of the time top versus bottom decile. To minimize look-ahead bias, we further add the assumption that annual COMPUSTAT variables are not available until 4 months after the fiscal year end.

We also improve on risk management. The returns of replicated factors are often dollar neutral but may still have market exposure. We measure market exposure by running a 36-months rolling regression of portfolio returns on market returns. We use the resulting rolling beta of the portfolio to dynamically hedge the raw portfolio returns. The key difference with the published methodology is the causal nature of our beta-hedge.

We report the distribution of Sharpe ratios and $t$-stats of our 72 factors in Table \ref{tab:stats_zoo}. We focus on in-sample results, \ie{} we take exactly the same period as in the paper. Using the same data and same predictors, but slightly different hedging procedure, we are able to replicate reasonably strong performance for our 72 factors. The mean Sharpe is at 0.98. Some factors do not replicate very well: Sharpe ratio at the first quartile breakpoint is only 0.43 with the corresponding $t$-stat of 1.89, a level of significance low by modern standard but not uncommon in the older literature. 

\begin{table}
    \caption{Summary statistics on our sample of 72 factors on their in-sample period.
    Annualised Sharpe ratio and $t$-stats were computed with monthly returns.
    All factors are beta-neutralised, computed on CRSP from 1963 to 2014.}
    \label{tab:stats_zoo}
    \centering
    \ra{1.3}
\begin{tabular}{@{} l *{2}{d{-1}} l *{2}{d{-1}} l *{2}{d{-1}} @{}}
\toprule
{} & \multicolumn{2}{c}{\textbf{Entire sample}} & \phantom{abc} & \multicolumn{2}{c}{\textbf{In-sample Sharpe $\mathbf{>0}$}} & \phantom{abc} & \multicolumn{2}{c}{\textbf{In-sample Sharpe $\mathbf{>0.3}$}} \\
\cmidrule{2-3} \cmidrule{5-6} \cmidrule{8-9}
{} &  \mc{Sharpe ratio} & \mc{$t$-stat} &&          \mc{Sharpe ratio} & \mc{$t$-stat} &&            \mc{Sharpe ratio} & \mc{$t$-stat} \\
\midrule
Mean   &          0.98 &   4.52 &&                  1.02 &   4.73 &&                    1.15 &   5.34 \\
Median &          0.85 &   3.79 &&                  0.88 &   3.84 &&                    0.99 &   4.55 \\
$Q_1$  &          0.43 &   1.89 &&                  0.46 &   1.98 &&                    0.69 &   2.75 \\
$Q_3$  &          1.33 &   6.24 &&                  1.38 &   6.39 &&                    1.46 &   7.10 \\
$N$    &         \multicolumn{2}{c}{$72$} &&                 \multicolumn{2}{c}{$69$} &&                   \multicolumn{2}{c}{$60$} \\
\bottomrule
\end{tabular}
\end{table}

Table \ref{tab:stats_zoo} also shows that a few factors fail to replicate at all. Three factors have a negative Sharpe ratio and another 9 have a Sharpe ratio which is less than 0.3. There are several reasons why we may fail to replicate the original results (\cite{mclean2016} face similar issues). First some authors may not have disclosed all the details and steps of the factor construction. Sometimes, an innocuous transformation of data (for instance, winsorization) may have an important impact on the final PnL. Second, the portfolio construction may contain a hidden look-ahead bias. We embargo the accounting data for 4 months after the end of fiscal year end, which is often more conservative than what is done in the literature. Last, in a few cases, we did not have access to all the data fields used by the authors, so we had to reconstruct the missing data using accounting identities. While the reconstructed quantities are typically very close to what the authors intended to use, they may alter the final PnL. 

We restrict our set of factors to strategies for which the in-sample Sharpe is at least 0.3, which is close to the 1.5 threshold on the $t$-stat that \cite{mclean2016} use to discard non-reproducible factors. This additional cut reduces the sample size somewhat but ensures we are focusing on factors that can reproduced.
\section{Out-of-sample Performance Drop}

In this section we look at the drop in performance of factors outside the training sample. We will explore three out-of-sample contexts: post-publication, large US stocks (which are a subset of the testing sample) and international pools. Here, we just focus on measures of performance drop and we delegate the study of its determinants to the next section.

\subsection{Measuring out-of-sample performance Loss}
\label{definition}

To measure the decline in the out-of-sample performance, we define a discount ratio as:
\begin{equation}\label{eq:dr}
  q\ :=\ \frac{SR_{OOS}}{SR_{IS}}\,,
\end{equation}
where $SR$ stands for Sharpe ratio, computed in-sample or out-of-sample. The discount ratio measures the \% drop in performance out-of-sample.

This measure is intuitive but faces an obvious challenge when the number of stocks varies between IS and OOS. This is because the Sharpe ratio of a factor is mechanically increasing with the number of stocks, a benefit of diversification. If OOS the number of stocks is smaller, even in the absence of actual performance decay, the measured Sharpe ratio will be lower. We thus need to adjust our Sharpe ratios for the number of stocks. 

In order to adjust for differences in size between IS and OOS, we adjust the Sharpe ratio by the number of stocks in the sample. To work out this adjustment, we need to find a formula that connects the Sharpe ratio with the number of stocks. To build this formula, start from the following model of the vector of returns:
\begin{equation}
    r_{t+1}\ =\ (b+\eta_{t+1})s_{t}\ +\ \beta R^m_{t+1}\ +\ \epsilon_{t+1}\,,
\end{equation}
where $s_{t}$ is a characteristic vector that can be measured. $\eta_{t+1}$ is an i.i.d. scalar shock. Stock returns are also exposed to the market via the vector of loadings $\beta$. In order to streamline algebra, we assume the elements of $s_t$ are ranks uniformly distributed between $-.5$ to $+.5$. 

Investors construct a long-short portfolio using $s_{t}$ as portfolio weights, $r^p_{t+1}= s_t'r_{t+1}$. We neglect market exposure here and assume that signals are orthogonal to $\beta$'s---which removes the need to hedge portfolio returns against the market. The Sharpe ratio associated with such portfolio is given by:
\begin{equation}
    \begin{split}
        SR\ &=\ \frac{b}{\sigma_\eta}\frac{1}{\sqrt{1+\frac{12\sigma_\epsilon^2}{\sigma_\eta^2 N}}} \\
        &\approx\ \underbrace{\frac{b}{\sigma_\eta}}_{SR_\infty}\left(1-\frac{6\sigma_\epsilon^2}{\sigma_\eta^2 }\frac{1}{N}\right)
    \end{split}
\end{equation}
where the factor 12 is simply $\int_{-1/2}^{1/2}{x^2 dx}$. The above formula makes it clear that the Sharpe ratio is an increasing function of the number of stocks traded. 

The above formula suggests two alternative ways of adjusting for $N$, depending on our prior on $\sigma_\eta$. 
\begin{itemize}
    \item \emph{Simple adjustment} If $\sigma_\eta=0$, there is no factor-specific risk, and the Sharpe-ratio writes as:
    \begin{equation}
        SR\ =\ \frac{b}{\sigma_\epsilon}\sqrt{\frac{N}{12}}\,,
    \end{equation}
    which suggests a simple adjustment: The size-adjusted Sharpe is the Sharpe ratio divided by $\sqrt{N}$. 
    \item \emph{Two-step adjustment} If $\sigma_\eta\neq0$, then the Sharpe is a linear function of $\frac{1}{N}$. We thus propose the following adjustment, for both IS and OOS Sharpes:
\begin{enumerate}
    \item For several values of $N$ we draw different subsamples of $N$ firms from the pool. Using each one of the subsamples we compute the Sharpe. Then, $SR(N)$ is the average across draws of these Sharpe ratios. 
    \item In the cross-section of $N$'s, regress $SR(N)$ on $\frac{1}{N}$, retrieve the constant term, which is the  size-adjusted Sharpe.
\end{enumerate}    
\end{itemize}

In most of our analyses, we will use the simple adjustment, but will also implement the two-step adjustment for international pools for robustness. 

\subsection{Post Publication Performance Drop}
\label{sec:prepub_postpub}

Our first analysis uses the post-publication period as out-of-sample context. In doing this, we follow the analysis of~\cite{mclean2016}. As discussed previously, and in line with~\cite{mclean2016}, we drop factors whose in-sample Sharpe is lower than 0.3.

\vspace{.2in}

\begin{figure}[tbph]
  \centering
  \begin{tikzpicture}
    \begin{axis}[
        grid=major,
        xlabel={Pre-publication Sharpe ratio},
        ylabel={Post-publication Sharpe ratio},
        legend entries={
          Observations,
          Mean value on 10 bins,
          $y=0.57\cdot x$,
          $y=x$},
        legend cell align={left},
        legend style={
          at={(0.02,0.98)},
          anchor=north west,
          font=\tiny,
        }
      ]
      \addplot+ [only marks, mark=o, lightgray] table {data/msr_mbn_is_postpub.dat};
      \addplot+ [
        only marks,
        mark=*,
        error bars/.cd,
        y dir=both,
        y explicit,
        ] table[x=x, y=y, y error=error] {data/bins_msr_mbn_is_postpub_error.dat};
      \addplot [blue, line width=1pt] table {data/reg_msr_mbn_is_postpub.dat};
      \addplot [black, dashed, line width=1pt] coordinates {(0,0) (3,3)};
    \end{axis}
  \end{tikzpicture}
  \begin{tikzpicture}
    \begin{axis}[
        grid=major,
        xlabel={Pre-publication Sharpe ratio},
        ylabel={Discount ratio},
        legend entries={
          Observations,
          $y=0.55$,
        },
        legend cell align={left},
        legend style={
          font=\tiny,
        }
      ]
      \addplot+ [only marks, mark=o] table {data/msr_mbn_is_discount.dat};
      \addplot [red, line width=1pt] table {data/reg_msr_mbn_is_discount.dat};
    \end{axis}
  \end{tikzpicture}
  \caption{Left: Sharpe ratios of factors, post-publication versus pre-publication.
  The regression is performed with OLS and has a $R^2$ of 67\%.
  Observations are cut into deciles and we compute the average (resp. median) Sharpe ratio in-sample (resp. post-publication).
  The error bars show the standard deviation in the $y$ direction.\\
  Right: Discount ratio versus pre-publication Sharpe ratio.
  The median discount ratio is 0.55.}
  \label{fig:prepub_postpub}
\end{figure}
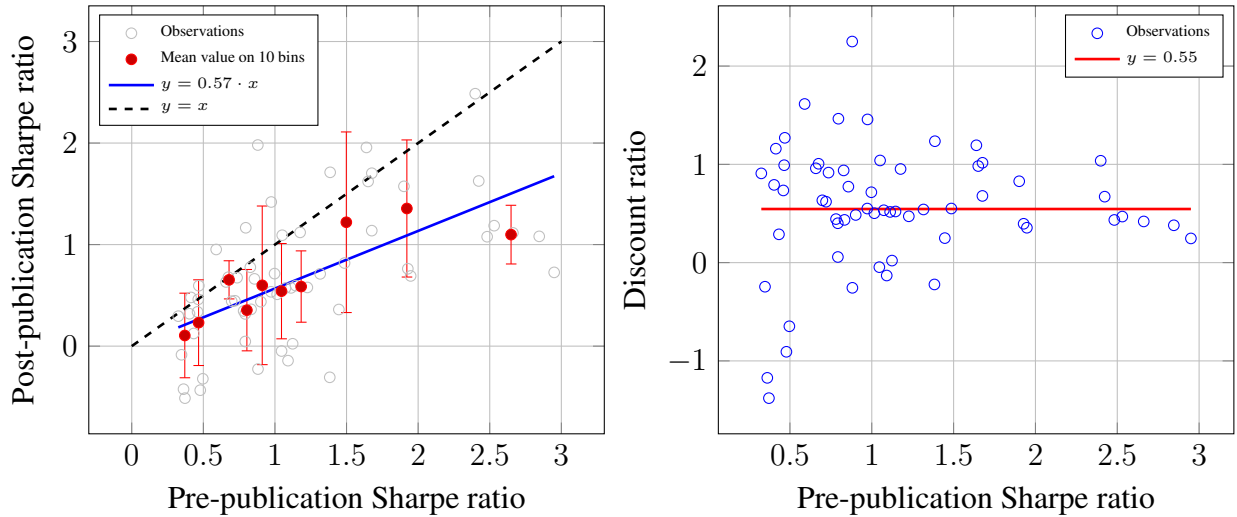

\bigskip

We describe the post-publication performance drop in Figure \ref{fig:prepub_postpub}. We plot in the left (resp. right) panel the Sharpe ratio in the post-publication period (resp. the discount ratio) as a function of the pre-publication Sharpe ratio. Consistently with \cite{mclean2016}, both charts confirm that, on average, post-publication Sharpe is below in-sample Sharpe. The binscatter plot in the left panel (in red), as well as the right panel, show that OOS Sharpe is well approximated by a linear function of IS Sharpe with slope .57. This slope means that, on average, post-publication, the Sharpe ratio drops by 43\%. 
This drop is slightly smaller than \cite{mclean2016}, who find a 58\% drop. Their set of factors is slightly larger than ours and we use a different hedging technique, but the magnitude is reassuringly similar. The approximate division by 2 of the Sharpe out-of-sample is consistent with the calibrated model of overfitting in \cite{rej2019}.

\subsection{Performance Drop on Liquid Stocks}
\label{sec:liquidstocksOOS}

We now test the performance drop of factors when restricting the pool of stocks. All of the replicated factors are defined on a subset of the CRSP universe where penny stocks and delistings are discarded, but it still includes many very small caps that cannot be traded at scale. Pools of larger stocks are relevant for asset managers looking for capacity, but not for academic research which was for long focused on rejecting non-predictability.

In this section our OOS test consists of looking at the performance in two of these larger pools: a pool composed of 1,000 most liquid stocks of CRSP and a one made of 500 most liquid stocks. The notion of ``out-of-sample'' test here is an abuse of language as these pools are actually a small subsample of the in-sample context.

\bigskip

\begin{figure}[htbp]
  \centering
  \begin{tikzpicture}
    \begin{axis}[
      grid=major,
      xlabel={Entire CRSP},
      ylabel={CRSP LQ},
      legend entries={
        {CRSP LQ 1,000, Sharpe ratio},
        {CRSP LQ 500, Sharpe ratio},
        {CRSP LQ 1,000, equivalent CRSP Sharpe},
        {CRSP LQ 500, equivalent CRSP Sharpe},
      },
      legend cell align={left},
      legend style={
        at={(0.02,0.98)},
        anchor=north west,
        font=\tiny,
      }
      ]
      \addplot+ [dashed, line width=1pt, mark=o, mark options={fill=blue}, blue] table {data/bins_msr_mbn_crsp_1000_error.dat};
      \addplot+ [dashed, line width=1pt, mark=o, mark options={fill=red}, red] table {data/bins_msr_mbn_crsp_500_error.dat};
      \addplot+ [solid, line width=1pt, mark=*, mark options={fill=blue}, blue] table {data/bins_msr_mbn_size_crsp_1000_error.dat};
      \addplot+ [solid, line width=1pt, mark=*, mark options={fill=red}, red] table {data/bins_msr_mbn_size_crsp_500_error.dat};
      \addplot [dashed, black] coordinates {(0,0) (2,2)};
    \end{axis}
    \end{tikzpicture}
  \caption{Raw and size-adjusted Sharpe ratios on liquidity pools vs. their original pool (as defined by the authors) counterparts. Observations are cut into 5 bins based on quantiles, we then compute the average (resp. median) Sharpe ratio on CRSP (resp. CRSP LQ 1,000 or 500).
  Factors are beta-neutralised and Sharpe ratios below 0.3 on the entire CRSP universe (in-sample) are excluded.}
  \label{fig:crsp_lq}
\end{figure}
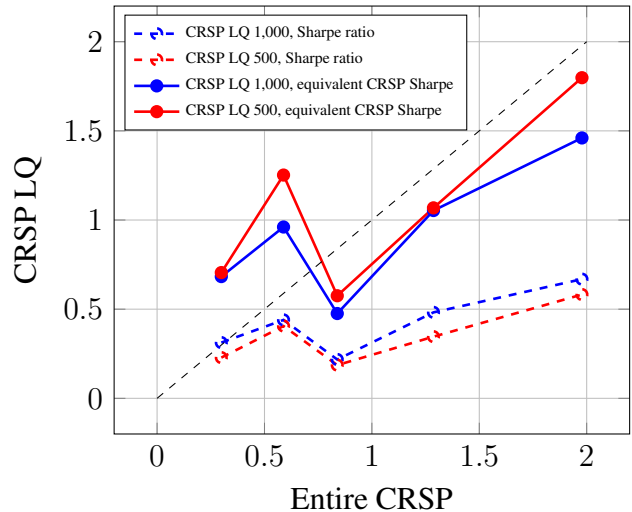

\bigskip

We plot in Figure \ref{fig:crsp_lq}  Sharpe ratios on liquidity pools against the Sharpe ratios on the original pool. The dashed lines correspond to the raw (not size adjusted) Sharpe ratios. Clearly, Sharpe ratios in larger pools are smaller than in-sample Sharpe ratios (the dashed lines are below the 45 degree line). A naive reading of this result would be that indeed, performance drops in the subsamples that have not been the main focus of researchers.

By design the pools of liquid stocks have a much smaller number of stocks than the in-sample context, so a size adjustment as described in section \ref{definition} is in order

\begin{equation}\label{eq:eq_sr}
    SR^*_{p}\ :=\ SR_p\sqrt{\frac{N_{CRSP}}{N_p}}\,,
\end{equation}
where $SR_p$ is the Sharpe ratio computed on pool $p$ and $N_p$ is the average number of stocks. Similarly, $N_{CRSP}$ is the average number of stocks in CRSP. Thus, $SR^*_{p}$ is the Sharpe ratio in pool $p$ adjusted for size. As discussed previously, the implicit assumption here is that Sharpe ratios are directly proportional to $\sqrt{N}$, \ie{} that predictability is a pure arbitrage, so that the Sharpe ratio would be infinite for $N \to \infty$. Another key assumption we are making here is that the number of stocks in each pool is constant and well approximated by the average number of stocks in the pool over the period. This assumption is a very rough approximation of reality for CRSP, where number of stocks peaked just before 2000 and declines ever since.

The solid lines of Figure \ref{fig:crsp_lq} report the result of this size adjusted approach. Now, the equivalent Sharpe ratio is much closer to the one computed on CRSP, even if on average still below the 45 degree line, hinting at some OOS drop in performance. We can quantify the average decay by regressing the size adjusted Sharpe on the CRSP sharpe, for each one of the two pools separately. We report regression results in appendix table \ref{tab:reg_crsp_lq_full}. We find that the OOS sharpe decay is 33\% for the largest 1,000 stocks, and 20\% for the largest 500.

\subsection{International pools}

We now analyze the performance drop on non-US pools of stocks. The rationale here is that is that most published factors were optimized on U.S. data, but were not tested on foreign stocks---a reason for this is the absence of stock returns on WRDS, the platform most finance academics use. Several papers (see for instance \cite{griffin2002}, \cite{asness2013}, \cite{RobStambaugh}) have looked at the robustness of a small number of well-known factors outside of the U.S., but papers investigating large set of factors (like \cite{mclean2016}, \cite{hou2018}) typically focus on U.S. data. 

We explore here how well our 72 factors do on 8 pools of stocks: China, Hong-Kong, Korea, Japan, Australia, Continental Europe, UK and Canada. To these international indices, we add the Russel 1,000 and Russel 3,000, which are broad indices of U.S. stocks. These pools correspond to well-known, liquid, stock indices, popular among large global investors. The period we study is shorter than that for U.S data: we focus on 2002-2018, because of limited historical data.

In order to study the drop in performance on these pools a size adjustment is necessary, see Section \ref{definition}. Figure \ref{fig:size} shows the variation over time of the number of stocks in each one of these pools. Most of these indices---except the China SE Shang Composite---have a reasonably stable number of stocks. Also, all of them are smaller than CRSP, some of them much smaller (the Australian pool has only 200 stocks) thus underlining the importance of the size adjustment .

\begin{table}[tpbh]
    \centering
    \ra{1.3}
        \caption{Median discount ratio.}
    \label{tab:median_dr_idx}
        \begin{tabular}{@{} ll *{4}{d{-1}} @{}}
        \toprule
\textbf{Country} & \textbf{Index/Pool} &  \mc{\textbf{Average size}} &    \mc{\textbf{Raw}} & \multicolumn{2}{c}{\textbf{Size adjusted}} \\
\cmidrule{5-6}
& & & & \mc{\textbf{Simple}} & \mc{\textbf{Complex}} \\
\midrule
\multicolumn{2}{l}{\emph{Panel A: US pools}} \\
United-States & CRSP post-publication       &        4694 &  0.55 &               0.52 &               0.54 \\
United States &  Russell 3000 Index     &        2933 &  0.20 &               0.26 &               0.22 \\
United States &        Russell 1000 Index   &        975 &  0.17 &               0.37 &               0.32 \\
\\
\multicolumn{2}{l}{\emph{Panel B: Non-US pools}} \\
Australia &         S\&P/ASX 200 Index   &         200 &  0.08 &               0.38 &               1.00 \\
 Europe &    Bloomberg European 500  &         505 &  0.09 &               0.24 &               0.37 \\
  Hong Kong &  Hang Seng Composite Index   &         312 &  0.20 &               0.75 &               0.74 \\
 Korea &         Korea Kospi Index  &         734 &  0.20 &               0.52 &               0.44 \\
 China &  China SE Shang Composite &        918 &  0.18 &               0.39 &               0.32 \\
 Canada &   S\&P/TSX Composite Index &        244 &  0.11 &               0.47 &               0.87 \\
Japan &     TOPIX 500 Index (TSE)     &    500 & -0.01 &              -0.02 &               0.32 \\
 United Kingdom &            FTSE 100 Index    &         100 &  0.08 &               0.53 &               1.78 \\
 \\
\multicolumn{2}{l}{\emph{Average non-US}} & 439 & 0.11 & 0.41 & 0.73 \\
\bottomrule
        \end{tabular}    
\end{table}

Table \ref{tab:median_dr_idx} reports the results. Each line corresponds to a pool of stocks. We show the discount ratios, which are the Sharpe ratios in each pools divided by the in-sample CRSP Sharpe ratio. These out-of-sample Sharpes are used raw (column 4) and using the ``simple'' size adjustment used in section \ref{sec:liquidstocksOOS}. Given that all pools are much smaller than the U.S., the simple size adjustment makes a big difference. Across the 8 international pools, the discount ratio is close to $.4$, a number similar to what is obtain by comparing pre- and post-publication Sharpe on CRSP. Thus, the Sharpe decay is significant but not full for our set of 72 factors. Such a decay would be consistent with overfitting: researchers testing many hypotheses on U.S. data only. It is also consistent with arbitrage, as for many of  factors in our zoo the period under consideration falls after the publication date. Quantitative investors would have had time to learn about the strategy on the U.S., and implemented it globally. Third possibility is that some stock anomalies reflect US-specific mispricing (due to differences in economics forces or accounting rules for instance). The rest of the paper tries to distinguish between these possibilities.
\section{Predicting out-of-sample performance decay}

The previous section has established that, whether we look at post-publication data, liquid U.S. pools or international pools, we observe a sizable drop in Sharpe ratio. The question we address now is whether the performance drop may be attributed to overfitting or arbitrage.

We focus here on comparing pre- and post-publication performance on CRSP. We construct in-sample covariates of the discount ratio related to either overfitting or arbitrage hypotheses. While we find that several variables succeed in explaining part of the discount ratios' variance, we are cognizant of the fact that we might have missed important (statistically-significant) proxies in both sets and thus the exact relative importance of one set w.r.t. the other may get reshuffled. 

\subsection{Arbitrage covariates}

We define four variables that are designed to test the potential influx of arbitrage capital into a factor once it is published. All of these variables are defined such that the expected sign of their effect on the discount ratio is negative (they indicate a larger scope for arbitrage capital inflows):
\begin{itemize}
\item \textbf{Holding period}.
The longer the holding period of a given factor, the cheaper it is to trade it, and the higher is its capacity \citep{bonelli}. We thus expect that slower strategies attract more arbitrage capital will be invested into the factor and their performance will decay more relative to their faster peers. 
Let us define the matrix $\bm{\omega}=\left(\omega_{t,s}\right)_{\substack{0<t\le T\\0<s\le N}}$ of weights, used to compute some factor $f$. We compute the holding period of stock $s$ as
\begin{equation}
    \mathcal{H}_s^f\ :=\ 2\cdot \frac{\sqrt{\langle\bm\omega_s^2\rangle-\langle\bm\omega_s\rangle^2}}{\left\langle\left|\Delta\bm\omega_s\right|\right\rangle}\,,
\end{equation}
where $\Delta\bm\omega_s:=\left(\omega_{t+1,s}-\omega_{t,s}\right)_{0<t<T}$ and $\langle\cdot\rangle$ denotes the mean operator.
We define the holding period of factor $f$ as the cross-sectional median: $\mathcal{H}^f:=M(\mathcal{H}^f_s)$.

\item \textbf{Amihud's liquidity}. As arbitrage capital moves in, we expect it to have a preference for liquid stocks. If liquidity correlates with factor exposure, arbitrage will hurt out-of-sample performance more and reduce the discount rate. To capture liquidity we use \emph{minus} Amihud's definition of illiquidity \citep{amihud2002}.
The liquidity of stock $s$ at day $t$ is
\begin{equation}
    L_{s,t}\ :=\ -\frac{|r_{s,t}|}{V_{s,t}}\,,
\end{equation}
with $r$ the (daily) return and $V$ the (daily) volume. We then compute the weighted average liquidity in the portfolio associated with factor $f$ and compare it to the average pool liquidity:
\begin{equation}
    \mathcal{L}^f_t\ :=\ \frac{1}{2}\left(\frac{\sum_{\omega_{s,t}>0}\omega_{s,t}L_{s,t}}{\sum_{\omega_{s,t}>0}\omega_{s,t}}+\frac{\sum_{\omega_{s,t}<0}\omega_{s,t}L_{s,t}}{\sum_{\omega_{s,t}<0}\omega_{s,t}}\right)\cdot\frac{N}{\sum_sL_{s,t}}.
\end{equation}
The liquidity of a factor is its time-series median $\mathcal{L}^f:=M(\mathcal{L}^f_t)$.

\item \textbf{Portfolio market cap to average market cap ratio}. This variable is a variant of the previous one, where liquidity is now proxied by firm size. If factor loadings are correlated with market capitalization, arbitrage capital will find it easier to move in and correct the mispricing. We construct this variable as the (weighted) average market cap in the long and short leg compared to the average market cap of all stocks in CRSP.
With $E_{s,t}$ the market equity of stock $s$ at time $t$, let us define
\begin{equation}
    \mathcal{M}^f_t\ :=\ \frac{1}{2}\left(\frac{\sum_{\omega_{s,t}>0}\omega_{s,t}E_{s,t}}{\sum_{\omega_{s,t}>0}\omega_{s,t}}+\frac{\sum_{\omega_{s,t}<0}\omega_{s,t}E_{s,t}}{\sum_{\omega_{s,t}<0}\omega_{s,t}}\right)\cdot\frac{N}{\sum_sE_{s,t}}.
\end{equation}
We take the time-series median as the final measure $\mathcal{M}^f:=M(\mathcal{M}^f_t)$.
Hence, large values of this proxy suggest an exposure to large, liquid, caps and we expect a lower discount ratio.

\item \textbf{Short leg market cap ratio}. This variable is a variant of the previous one design to capture the hard-to-borrow effect. Small caps may be notoriously difficult to short and thus our proxy should be a proxy for the hard-to-borrow effect. We define this variable as the (weighted) average market cap in the short leg relative to the average market cap of the market. If the ratio is large, the short leg is cheap to implement and arbitrage capital is expected to move in aggressively.
\end{itemize}

\begin{table}[tbph]
    \caption{Stats for arbitrage variables before standardization. We only retain strategies with in-sample Sharpe ratio on CRSP higher than 0.3.}
    \centering
    \ra{1.3}
\begin{tabular}{@{} l *{5}{d{-1}} @{}}
\toprule
{} &  \mc{\textbf{Mean}} &  \mc{\textbf{Median}} &  \mc{$\mathbf{Q_1}$} &  \mc{$\mathbf{Q_3}$} &   \mathbf{N} \\
\midrule
log holding period     &  6.08 &    6.22 &   5.91 &   6.40 &  58 \\
log mkt cap long short &  0.88 &    0.62 &  -0.08 &   2.00 &  58 \\
log mkt cap short      &  0.95 &    0.66 &   0.04 &   1.64 &  58 \\
liquidity              & -0.64 &   -0.74 &  -0.99 &  -0.17 &  58 \\
\bottomrule
\end{tabular}
\end{table}

\subsection{Overfitting covariates}

We now propose different proxies that may be relevant to multiple testing hypothesis. We use the same convention as for the arbitrage variables: a larger value of each variable constructed should translate to \emph{higher chance} of overfitting.

We start with a variable that proxies for the fact that models may have been ``selected'' to fit the data in-sample:
\begin{itemize}
\item \textbf{Low $t$-stat}.~\cite{harvey2017} claims, drawing upon Bonferroni inequalities, that statistical significance in investment research requires a $t$-stat in excess of 3 in order to counteract the problem of multiple testing.  Thus, we construct a dummy variable equal to 1 if the $t$-stat of the factor is less than 3. When this is the case, the risk of multiple hypothesis testing is higher, and the expected discount ratio should be lower.
\end{itemize}

\bigskip

If there are many ways of testing the hypothesis then, conditional on publication, we expect a larger drop of performance out-of-sample. We thus propose four variables that seek to capture the  ``flexibility'' of the sorting variable in achieving different in-sample Sharpe ratios:
\begin{itemize}
\item \textbf{Log quantile span}. 
We define a family of strategies for each given factor by varying the top and bottom quantile used to define the long and short legs of the portfolio. For each quantile  $q \in \{5\%, 10\%, 15\%, 20\%, 25\%, 30\%, 35\%\}$, we compute one in-sample Sharpe $SR_q$. We then define this flexibility variable of factor $f$ as
\begin{equation}
    \mathcal{F}^f\ :=\ \ln\frac{\sqrt{\langle SR^2_q\rangle-\langle SR_q\rangle^2}}{SR_{IS}}.
\end{equation}
The intuition is that the larger this variable is, the more sensitive is the factor's performance \wrt{} portfolio construction and the less likely it is to generalize outside the training period.

\item \textbf{Deviation from best q}. This is a variation on the log quantile span theme. Instead of log of the range, we look at the quantile $q^*$ that maximizes the in-sample Sharpe. We then compute the difference between  $SR_{q^*}$ and the in-sample Sharpe in the baseline specification. If this difference is large, it means the researcher has a lot of scope to come up with the ``best-looking'' strategy. 
\item \textbf{Dummy number of COMPUSTAT fields}. We define this dummy variable as 1 if the number of COMPUSTAT fields used to define the factor is more than two. The intuition behind this variable is that, once researchers start combining multiple COMPUSTAT fields, the number of possible combinations explodes and some of them will lead to false positive signals in-sample.
Let us point out though, that sometimes authors use several Compustat fields in order to define certain well-known financial / accounting metrics not available as fields in the Compustat.
This weakens our argument only slightly, as a raft of accounting metrics exists thus considerably extending the universe of possible false positives. 

\item \textbf{Dummy number of operations}.
This dummy variable is 1 if the number of operations used to define the factor is more than two. The intuition is similar to the number of COMPUSTAT fields. More operations may be a sign that the sorting variable is only useful when ``properly groomed''.
\end{itemize}    

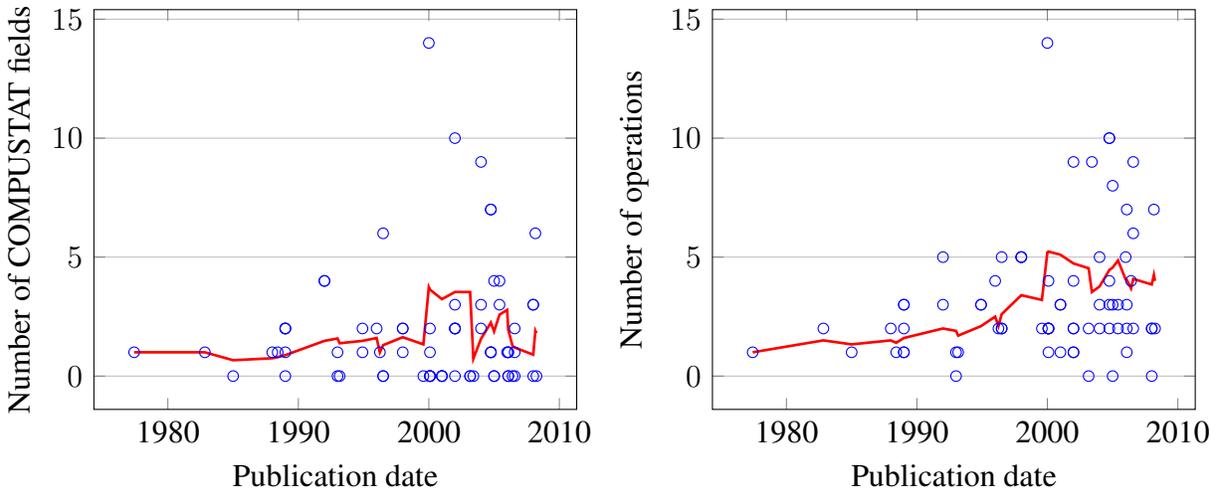
\begin{figure}[tbph!]
    \centering
    \begin{tikzpicture}
        \begin{axis}[
            width=8cm,
            date coordinates in=x,
            xtick={1970-01-01, 1980-01-01, 1990-01-01, 2000-01-01, 2010-01-01}, 
            xticklabel=\year,
            ymajorgrids,
            xlabel={Publication date},
            ylabel={Number of COMPUSTAT fields},
            ]
            \addplot+ [blue, only marks, mark=o] table {data/nb_fields_dapublished_crsp.dat};            
            \addplot [red, line width=1pt] table {data/nb_fields_dapublished_crsp_rolling.dat};
        \end{axis}
    \end{tikzpicture}
    \begin{tikzpicture}
        \begin{axis}[
            width=8cm,
            date coordinates in=x,
            xtick={1970-01-01, 1980-01-01, 1990-01-01, 2000-01-01, 2010-01-01}, 
            xticklabel=\year,
            ymajorgrids,
            xlabel={Publication date},
            ylabel={Number of operations},
            ]
            \addplot+ [blue, only marks, mark=o] table {data/nb_operations_dapublished_crsp.dat};            
            \addplot [red, line width=1pt] table {data/nb_operations_dapublished_crsp_rolling.dat};
        \end{axis}
    \end{tikzpicture}
    \caption{Number of Compustat fields (resp. operations) used to compute the factors' characteristic, as a function of publication date.
    We only show factors whose in-sample Sharpe ratio is greater than 0.3.
    One dot per factor.
    The red line draws a moving average of size 5 (factors).}
    \label{fig:fields_operations_dapublished}
\end{figure}

In Figure \ref{fig:fields_operations_dapublished}, we show that there is a tendency in exploring more and more complex predictors in recent research. In this Figure, we plot the number of fields and the number of operations, for each publication, as a function of the publication date of the predictor. To smooth out the raw data, we also show a red line representing the 5-factor moving average (average for the last 5 factors published). These two charts describe the increasing complexity of asset-pricing research for predictability. Such complexity may just reflect that more complex formulas are needed to find new signals, or plain overfitting. Our cross-sectional analysis below investigates this.

\bigskip

The last set of three variables indicate the small-sample sensitivity of the in-sample Sharpe estimate.
\begin{itemize}
\item \textbf{Minus number of months in-sample}. This variable is yet another way of capturing the risk of overfitting. Back tests over long periods are more likely to pick up ``stationary'' anomalies and not only transient market inefficiencies. Such persistent anomalies should have a higher chance of working outside the training period. Even if all stock phenomena were to be stationary, it is easier to ``mine'' statistically significant relationships on a short sample. Thus, if the number of months in-sample is smaller (the variable is larger), we expect the discount ratio to be smaller. 

\item \textbf{Log of subset std}. This variable tries to capture the robustness of the factor strategy to removing a small subset of stocks. To compute this, we randomly draw a 10\% subset of stocks every period. We drop this subset and calculate the Sharpe ratio on the remaining 90\% of the sample. We repeat this step 100 times and compute the standard deviation of these Sharpe ratio. This variable captures the sensitivity of a factor's performance to removing a small number of observations. If this sensitivity is high, the in-sample Sharpe is more likely to depend on a small subset of stocks that may exhibit different properties out-of-sample or disappear from the pool altogether. 

\item \textbf{Diff SR with dropped data}. This second variable seeks to directly capture the contribution of outliers to the in-sample Sharpe ratio. Such outliers are potentially important since returns are known to be non-normal (see e.g. \cite{gabaix2016}). To appraise their contribution to the in-sample Sharpe ratio, we implement the methodology proposed by \cite{broderick2020}. We define a new version of factors, which uses all of the in-sample observations, save for the 0.1\% observations that are most impactful. Following the definition of \cite{broderick2020}, these observations are the top 0.1\% portfolio returns\footnote{More precisely, each stock contributes $\omega_{s,t} r_{s,t+1}$ to the strategy P\&L every month. We drop the top 0.1\% contributions.} every month. Since our factors are defined in a long/short way, these may came either from the long or the short leg.  Once this has been done, we recompute the in-sample Sharpe. The difference between this new Sharpe and the full sample in-sample Sharpe is our variable. The larger it is, the more in-sample performance is sensitive to a few stocks and the more likely performance is to drop out of sample. 
\end{itemize}

To make these variables comparable to one another, we standardize them. Thus, by construction, all variables have a mean of zero and a standard deviation of 1.

\begin{table}[tbph]
    \caption{Stats for overfitting variables before standardization. We only retain strategies with in-sample Sharpe ratio on CRSP higher than 0.3.}
    \centering
    \ra{1.3}
\begin{tabular}{@{} l *{5}{d{-1}} @{}}
\toprule
{} &  \mc{\textbf{Mean}} &  \mc{\textbf{Median}} &  \mc{$\mathbf{Q_1}$} &  \mc{$\mathbf{Q_3}$} &   \mathbf{N} \\
\midrule
dumy tstat         &   0.28 &    0.00 &   0.00 &   1.00 &  60 \\
log q span cv      &  -1.95 &   -2.11 &  -2.43 &  -1.37 &  58 \\
diff from best q   &   0.15 &    0.08 &   0.03 &   0.21 &  58 \\
dumy nb fields     &   0.35 &    0.00 &   0.00 &   1.00 &  60 \\
dumy nb operations &   0.75 &    1.00 &   0.75 &   1.00 &  60 \\
sqrt nb months is  & -16.06 &  -16.57 & -18.52 & -12.64 &  60 \\
log std subset     &  -3.41 &   -3.45 &  -3.82 &  -2.91 &  57 \\
diff w drop data   &   5.42 &    5.35 &   4.34 &   6.42 &  58 \\
\bottomrule
\end{tabular}
\end{table}

\subsection{Publication Date}

Our 11$^{th}$ variable, which is a proxy for both arbitrage and overfitting. The date of publication---measured in Unix time, \ie{} the number of seconds elapsed since Jan 1st 1970.
We expect the publication year to negatively affect discount ratios. It is reasonable to believe that number of genuine factors is rather limited. As the ``true'' factors are revealed by researchers, people look harder and harder for false positives.
The negative effect of the publication year on discount ratio is also consistent with the arbitrage view. Given the rise of the hedge fund industry over the past few decades, a reasonable conjecture would be that arbitrage capital is more likely to ``rush in'' after publication than before. 

\bigskip

\begin{figure}[tbph]
  \centering
    \begin{tikzpicture}
        \begin{axis}[
            width=7cm,
            date coordinates in=x,
            xtick={1970-01-01, 1980-01-01, 1990-01-01, 2000-01-01, 2010-01-01}, 
            xticklabel=\year,
            ymajorgrids,
            xlabel={Publication date},
            ylabel={Discount ratio},
            legend entries={
              Observations,
              $y=2.03-1.60\times10^{-9}\cdot x$,
            },
            legend cell align={left},
            legend style={
              font=\tiny,
            },
        ]
            \addplot+ [blue, only marks, mark=o] table {data/discount_dapublished_crsp.dat};            
            \addplot [red, line width=1pt] table {data/discount_dapublished_crsp_reg.dat};
        \end{axis}
    \end{tikzpicture}
      \begin{tikzpicture}
    \begin{axis}[
            width=7cm,
        grid=major,
        xlabel={Number of days around publication date},
        ylabel={Log returns}
      ]
      \addplot [blue, line width=1pt] table {data/pnl_kink_dapublished.dat};
    \end{axis}
  \end{tikzpicture}
    \caption{Left Panel: Discount ratio as a function of publication date.
    Factors are beta-neutralised and computed on CRSP, conditional on its in-sample Sharpe ratio being greater than 0.3.
    One dot per factor.
    The red line draws a linear trend, fitted on blue dots, where dates are represented in Unix time.
    Right Panel: Average PnL of factors, synced on publication date.
  The average PnL is detrended based on the pre-publication period.
  Factors are beta-neutralised and, in order to compare them, risk-managed.
  Computed on CRSP, conditional on its in-sample Sharpe ratio being greater than 0.3.
  }
    \label{fig:timetime}
\end{figure}
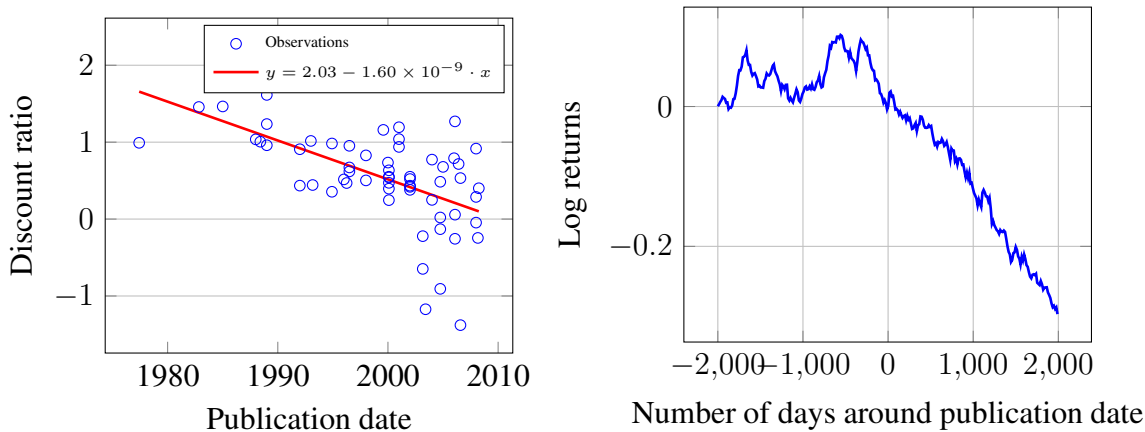

Consistent with both of these views, the left Panel of Figure \ref{fig:timetime} shows that Sharpe decay after publication has increased substantially over time, especially since 2000.%
\footnote{This comes from the decay of out-of-sample Sharpe ratio, while the in-sample Sharpe stays flat on average with publication date. See figure \ref{fig:is_oos_dapublished}.}
This movement was quick, with the discount ratio decreasing by 0.05 every year, \ie{} the expected haircut increases by 5ppt of the in-sample Sharpe every year. On the right panel of Figure \ref{fig:timetime}, we show the average detrended PNL of strategies aroung the date of publication, going from 2,000 days before to 2,000 days after. For each strategy, the PNL is calculated as the cumulative return of investing one dollar in the risk-controlled strategy. The detrended PNL is computed after taking out the average slope in the pre-publication period. The right Panel clearly shows a very fast performance decay after publication (actually, about one year and a half before publication, which is consistent with pre-print circulation). This suggests that performance decays extremely fast, possibly too fast to be explained by arbitrage capital inflows.

\subsection{Regression Results}

In order to minimize the probability of false positives or false negatives, we start with univariate regressions results. This helps to identify relevant variables and avoids the pitfalls of collinearity. 

In table \ref{tab:corr_arbitrage} we present the correlations between arbitrage variables. We see that three of our covariates: log mkt cap long short, log mkt cap short and illiquidity, are highly correlated. This is to be expected as they all proxy for effects that are related to liquidity. 

\bigskip

\begin{table}[tbph]
    \caption{OLS univariate regressions of discount ratio on arbitrage variables. $t$-stats in parentheses.}
    \label{tab:full_reg_arbitrage}
    \centering
    \ra{1.3}
\begin{tabular}{@{} l *{4}{d{5}} @{}}
\toprule
{} &      1 &      2 &      3 &      4 \\
\midrule
const                  &   0.87 &   0.70^{***} &   0.72^{***} &   0.36^{***} \\
                       &  (1.22) &   (7.8) &  (7.86) &  (2.64) \\
log holding period     &  -0.05 &        &        &        \\
                       & (-0.41) &         &         &         \\
log mkt cap long short &        &  -0.14^{**} &        &        \\
                       &         & (-2.48) &         &         \\
log mkt cap short      &        &        &  -0.15^{***} &        \\
                       &         &         & (-2.67) &         \\
liquidity              &        &        &        &  -0.34^{*} \\
					   &         &         &         & (-1.95) \\
\midrule
$N$                      &  58 &  58 &  58 &  58 \\
$R^2$                     &   0.00 &   0.10 &   0.11 &   0.06 \\
\bottomrule
\multicolumn{5}{l}{$^{***}\ p<0.01$; $^{**}\ p<0.05$; $^{*}\ p<0.1$}\\
\end{tabular}
\end{table}

Table \ref{tab:full_reg_arbitrage} reports the results of the univariate regressions on each  one of the liquidity proxies. All coefficients are negative, as expected. The market cap based coefficients are both significant, as is the Amihud-liquidity-based one. We conclude that liquidity constraints, as expected, prevent some of market inefficiencies from being completely arbitraged away.

\bigskip

We now move to overfitting variables. In Table \ref{tab:corr_overfitting} we present the correlations between overfitting variables. Much of the correlation structure does not come as a surprise and is pretty much mechanical like the correlation between \textit{dapublished} and \textit{sqrt nb months is}, or between \textit{dapublished} and \textit{number of operations}: with time authors explored more complex characteristics, see figure \ref{fig:fields_operations_dapublished}.

\begin{table}[tbph!]
    \caption{OLS univariate regressions of discount ratio on overfitting variables. $t$-stats in parentheses.}
    \label{tab:full_reg_overfitting}
    \centering
    \ra{1.3}
    \small
\begin{tabular}{@{} l *{9}{d{0.5}} @{}}
\toprule
{} &      1 &      2 &      3 &      4 &      5 &      6 &      7 &      8 &      9 \\
\midrule
const              &   0.59^{***} &   0.52^{***} &   0.54^{***} &   0.63^{***} &   0.92^{***} &   0.55^{***} &   0.59^{***} &   0.59^{***} &   0.53^{***} \\
              & (6.12) &  (6.09) &  (6.39) &  (6.29) &   (5.9) & (6.73) &  (8.01) &  (7.99) &  (7.72) \\
dumy tstat         &  -0.15 &        &        &        &        &        &        &        &        \\
         & (-0.8) &         &         &         &         &        &         &         &         \\
log q span cv      &        &  -0.10 &        &        &        &        &        &        &        \\
      &        & (-1.08) &         &         &         &        &         &         &         \\
diff from best q   &        &        &  -0.07 &        &        &        &        &        &        \\
   &        &         & (-0.83) &         &         &        &         &         &         \\
dumy nb fields     &        &        &        &  -0.23 &        &        &        &        &        \\
     &        &         &         & (-1.37) &         &        &         &         &         \\
dumy nb operations &        &        &        &        &  -0.49^{***} &        &        &        &        \\
 &        &         &         &         & (-2.72) &        &         &         &         \\
sqrt nb months is  &        &        &        &        &        &   0.08 &        &        &        \\
  &        &         &         &         &         & (0.91) &         &         &         \\
log std subset     &        &        &        &        &        &        &  -0.16^{**} &        &        \\
     &        &         &         &         &         &        & (-2.04) &         &         \\
diff w drop data   &        &        &        &        &        &        &        &  -0.21^{***} &        \\
   &        &         &         &         &         &        &         & (-2.99) &         \\
dapublished        &        &        &        &        &        &        &        &        &  -0.35^{***} \\
        &        &         &         &         &         &        &         &         & (-5.06) \\
\midrule
$N$                  &  60 &  58 &  58 &  60 &  60 &  60 &  57 &  58 &  60 \\
$R^2$                 &   0.01 &   0.02 &   0.01 &   0.03 &   0.11 &   0.01 &   0.07 &   0.14 &   0.31 \\
\bottomrule
\multicolumn{10}{l}{$^{***}\ p<0.01$; $^{**}\ p<0.05$; $^{*}\ p<0.1$}\\
\end{tabular}
\end{table}

The regression results are provided in Table \ref{tab:full_reg_overfitting}. The statistically significant overfitting variables capture different overfitting-related effects, as it may be inferred from lower level of correlations between them. First, publication date emerges as a very strong predictor of OOS Sharpe decay. As seen in figure \ref{fig:timetime}, factors tend to be more overfitted when published more recently. Second, among the ``flexibility'' variables, the number of operations is the only one that comes out significant, although, as expected, all are negative.
Finally, the two measures that seek to capture the dependence on a small number of observations do well at forecasting lower discount ratios. In-sample factors that derive a significant part of their performance from a small subset of stocks do not generalize well. Also, our measure of sensitivity of in-sample Sharpe to a random 10\% of the sample (log std subset) also strongly predicts performance decay.

Overall, date of publication, the sensitivity to outliers suggested by \cite{broderick2020} and the number of operations have the largest univariate $R^2$, \ie{} .30, .14 and .11. Put together, the 3 variables have an $R^2$ of 0.36, essentially driven by the first two variables. 

\subsection{Putting it all together}

Last, we run a horse race between arbitrage and overfitting variables. We construct two variables: ``arbitrage vulnerability'' and ``overfitting vulnerability'', by taking an arithmetic mean for each set (4 arbitrage and 6 overfitting variables, respectively). Remember that all variables have been standardized to an s.d. of 1 and a mean of zero, so taking a mean gives an equal weight to all variations. To these two summary variables, we add the publication date variable. 

\begin{table}[tbph]
    \centering
    \ra{1.3}
    \caption{Arbitrage vs overfitting Vulnerabilities}
    \label{tab:horserace}
    \small
\begin{tabular}{@{} l *{5}{d{-1}} @{}}
\toprule
Dependent variable: &\multicolumn{5}{c}{Discount Ratio} \\
&(1)&(2)&(3)&(4)&(5) \\

\midrule
Year of publication -1990&        -.05^{***}&               &               &       -.041^{***}&       -.044^{***}\\
                    &      (-4.9)   &               &               &      (-4.9)   &      (-5.2)   \\
Arbitrage vulnerability&               &        -.28^{**} &               &        -.13   &               \\
                    &               &      (-2.6)   &               &      (-1.5)   &               \\
Overfitting vulnerability&               &               &        -.34^{***}&        -.28^{***}&        -.31^{***}\\
                    &               &               &      (-3.1)   &        (-3.0)   &      (-3.5)   \\
Constant               &           1.0^{***}&         .58^{***}&         .55^{***}&         .92^{***}&         .94^{***}\\
                    &       (8.7)   &       (7.7)   &       (7.4)   &       (9.6)   &       (9.8)   \\
\midrule
$N$        &          60   &          58   &          55   &          55   &          55   \\
$R^2$               &        0.30   &        0.11   &        0.15   &        0.47   &        0.45   \\

\bottomrule
\multicolumn{6}{l}{$^{***}\ p<0.01$; $^{**}\ p<0.05$; $^{*}\ p<0.1$}\\
\end{tabular}
\end{table}

In Table \ref{tab:horserace}, we report the result of the predictive regression. First, the date of publication has a strong explanatory power on the cross-section of Sharpe decays, with an $R^2$ of 0.30. Second, the combined overfitting vulnerability variable is also quite strong, with an $R^2$ of .15. Arbitrage vulnerability has a weaker predictive power and a lower---yet significant---$t$-stat. When we put the three variables together, we end up with an $R^2$ of .47. Removing the arbitrage variable diminishes the $R^2$ only slightly suggesting its mariginal importance. 
\section{Conclusions}

In this article we have attempted to ``shake, rattle and roll'' a large zoo of academic investment factors in order to glean insights about their robustness. We reproduce the out-of-sample performance drop of on average 50\% that was reported by other authors. What makes our work original, however, is the attempt to explain this drop in a qualitative and quantitative way. 

To that effect we have proposed two hypothesis. The first one holds that the drop is a pure overfitting effect. In other words, the authors have fine-tuned the (on average) genuine effect to the data used for backtesting in order to make their findings more statistically significant. This fine-tuning went far beyond what the true effect was and thus only part of the performance is recovered out-of-sample. Second hypothesis posits that once a profitable strategy is disseminated, market participants move in in order to monetize the effect. This in turn weakens or even obliterates the effect going forward.

In order to be able to tell which (if any) of these two hypothesis is more likely, we have constructed a set of explanatory variables for each hypothesis and use univariate regressions to test each variable separately. We find statistically significant coefficients for several variables from both sets. This suggests that both forces are at work and the decay in performance is a joint effect of these forces.

We have also studied whether the factors generalize to different pools in different countries. This needs to be done with care, as the size of such pools is typically much lower than the pool on which the original research was carried out (CRSP). After adjusting for the pool size, we find that factors lose part of their performance when "uprooted" from their original "habitat". This may be because the effect these factor capture is specific to US market and should not be expected to hold outside US, or is another indication of the overfitting prowess of academicians. Since our price data on international pools starts only around 2000 we are unable to make a conclusive distinction between in-sample and out-of-sample performance outside US.

\newpage

\bibliographystyle{ecta}
\bibliography{zoo_bib}

\newpage

\appendix

\renewcommand{\thetable}{\Alph{section}\arabic{table}}
\renewcommand{\thefigure}{\Alph{section}\arabic{figure}}
\setcounter{table}{0}
\setcounter{figure}{0}

\setcounter{table}{0}
\renewcommand{\thetable}{A.\arabic{table}}
\renewcommand{\thefigure}{A.\arabic{figure}}    

\section{Appendix Tables}
\label{sec:a1}

\begin{table}[tbph]
\caption{List of factors and the corresponding references}
\label{tab:list_factor}
  \centering
  \ra{1.1}
  \begin{tabular}{@{}ll@{}}
    \toprule
    \textbf{Factor}                                           & \textbf{Reference article}\\
    \midrule
    Price earnings                                   & \cite{basu1977}\\
    Unexpected quarterly earnings                    & \cite{rendleman1982}\\
    Long term reversal                               & \cite{debondt1985}\\
    Debt equity                                      & \cite{bhandari1988}\\
    Change in inventory-to-assets                    & \cite{ou1989}\\
    Change in dividend per share                     & \cite{ou1989}\\
    Change in capital expenditures-to-assets         & \cite{ou1989}\\
    Return on total assets                           & \cite{ou1989}\\
    Debt repayment                                   & \cite{ou1989}\\
    Depreciation-to-PP\&E                            & \cite{holthausen1992}\\
    Change in depreciation-to-PP\&E                  & \cite{holthausen1992}\\
    Change in total assets                           & \cite{holthausen1992}\\
    Size                                             & \cite{fama1993}\\
    Book-to-market                                   & \cite{fama1993}\\
    Momentum                                         & \cite{jegadeesh1993}\\
    Sales growth                                     & \cite{lakonishok1994}\\
    Cash flow-to-price                               & \cite{lakonishok1994}\\
    Working capital accruals                         & \cite{Sloan1996}\\
    Sales-to-price                                   & \cite{barbee1996}\\
    Share turnover                                   & \cite{haugen1996}\\
    Cash flow-to-price variability                   & \cite{haugen1996}\\
    Trading volume trend                             & \cite{haugen1996}\\
    \bottomrule
    &\emph{\hspace{2cm} see next page}\\
  \end{tabular}
\end{table}

\newpage
\begin{table}[tbph]
\ContinuedFloat
\caption{List of factors and the corresponding references (continued)}
  \centering
  \ra{1.1}
  \begin{tabular}{@{}ll@{}}
    \toprule
    \textbf{Factor}                                           & \textbf{Reference article}\\
    \midrule
    Inventory                                        & \cite{abarbanell1998}\\
    Gross margin                                     & \cite{abarbanell1998}\\
    Capital expenditures                             & \cite{abarbanell1998}\\
    Industry momentum                                & \cite{moskowitz1999}\\
    F-score                                          & \cite{piotroski2000}\\
    Industry adjusted book-to-market                 & \cite{asness2000}\\
    Industry adjusted cash flow-to-price             & \cite{asness2000}\\
    Industry adjusted size                           & \cite{asness2000}\\
    Industry adjusted momentum                       & \cite{asness2000}\\
    Industry adjusted long term reversal             & \cite{asness2000}\\
    Industry adjusted short term reversal            & \cite{asness2000}\\
    Dollar volume                                    & \cite{chordia2001}\\
    Dollar volume coefficient of variation           & \cite{chordia2001}\\
    Share turnover coefficient of variation          & \cite{chordia2001}\\
    Change in inventory                              & \cite{thomas2002}\\
    Change in current assets                         & \cite{thomas2002}\\
    Depreciation                                     & \cite{thomas2002}\\
    Change in accounts receivable                    & \cite{thomas2002}\\
    Other accruals                                   & \cite{thomas2002}\\
    Illiquidity                                      & \cite{amihud2002}\\
    Liquidity                                        & \cite{pastor2003}\\
    Idiosyncratic return volatility x book-to-market & \cite{ali2003}\\
    Price x book-to-market                           & \cite{ali2003}\\
    \bottomrule
    &\emph{\hspace{2cm} see next page}\\
  \end{tabular}
\end{table}

\begin{table}[tbph]
\ContinuedFloat
\caption{List of factors and the corresponding references (continued)}
  \centering
  \ra{1.1}
  \begin{tabular}{@{}ll@{}}
    \toprule
    \textbf{Factor}                                           & \textbf{Reference article}\\
    \midrule
    Operating cash flow-to-price                     & \cite{desai2004}\\
    Abnormal corporate investment                    & \cite{titman2004}\\
    Accrual quality                                  & \cite{francis2004}\\
    Earnings persistence                             & \cite{francis2004}\\
    Smoothness                                       & \cite{francis2004}\\
    Value relevance                                  & \cite{francis2004}\\
    Timeliness                                       & \cite{francis2004}\\
    Tax income-to-book income                        & \cite{lev2004}\\
    Price delay                                      & \cite{hou2005}\\
    Firm age                                         & \cite{jiang2005}\\
    Duration                                         & \cite{jiang2005}\\
    Change in current operating assets               & \cite{richardson2005}\\
    Change in non-current operating liabilities      & \cite{richardson2005}\\
    Growth in capital expenditures                   & \cite{anderson2006}\\
    Growth in capital expenditures (alternative)     & \cite{anderson2006}\\
    Low volatility                                   & \cite{ang2006}\\
    Low beta $\Delta$VIX                             & \cite{ang2006}\\
    Zero trading days                                & \cite{liu2006}\\
    Composite issuance                               & \cite{daniel2006}\\
    Intangible return                                & \cite{daniel2006}\\
    Earnings surprises x revenue surprises           & \cite{jegadeesh2006}\\
    Industry concentration                           & \cite{hou2006}\\
    Change in shares outstanding                     & \cite{pontiff2008}\\
    Seasonality                                      & \cite{heston2008}\\
    Investment                                       & \cite{lyandres2008}\\
    Investment growth                                & \cite{xing2008}\\
    Change in asset turnover                         & \cite{soliman2008}\\
    \bottomrule
  \end{tabular}
\end{table}

\newpage
\begin{table}[tbph]
    \centering
    \ra{1.3}
    \caption{OLS regressions of equivalent Sharpe ratio of CRSP LQ 500 (resp. 1,000) on CRSP.
    $t$-stats in parenthesis.}
    \label{tab:reg_crsp_lq_full}
\begin{tabular}{@{} l *{2}{d{5}} @{}}
\toprule
{} &  \mc{CRSP LQ 500} &  \mc{CRSP LQ 1,000} \\
\midrule
coef &         0.80^{***} &           0.67^{***} \\
     &         (8.68) &           (8.52) \\
\midrule
$N$    &        60 &          60 \\
$R^2$   &         0.56 &           0.55 \\
\bottomrule
\multicolumn{3}{l}{$^{***}\ p<0.01$; $^{**}\ p<0.05$; $^{*}\ p<0.1$}\\
\end{tabular}
\end{table}

\newpage

\begin{table}[tbph]
    \caption{Correlation matrix for arbitrage variables.}
    \label{tab:corr_arbitrage}
    \centering
    \ra{1.3}
\begin{tabular}{@{} l *{4}{d{-1}} @{}}
\toprule
{} &  \mc{log holding period} &  \mc{log mkt cap long short} &  \mc{log mkt cap short} &  \mc{liquidity} \\
\midrule
log holding period     &                1.00 &                         &                    &            \\
log mkt cap long short &               -0.23 &                    1.00 &                    &            \\
log mkt cap short      &               -0.08 &                    0.89 &               1.00 &            \\
liquidity              &               -0.07 &                    0.78 &               0.62 &        1.00 \\
\bottomrule
\end{tabular}
\end{table}

\newpage

\begin{table}[tbph]
    \caption{Correlation matrix for overfitting variables: dts (dumy tstat), lqscv (log q span cv), dbq (diff from best q), dnf (dumy nb fields), dno (dumy nb operations), snm (sqrt nb months is), lss (log std subset), ddd (diff w drop data), dp (dapublished).}
    \label{tab:corr_overfitting}
    \centering
    \ra{1.3}
    \begin{tabular}{@{} l *{9}{d{-1}} @{}}
\toprule
{} &   \mc{dts} &  \mc{lqscv} &   \mc{dbq} &   \mc{dnf} &   \mc{dno} &   \mc{snm} &   \mc{lss} &   \mc{ddd} &   \mc{dp} \\
\midrule
dts   &  1.00 &        &       &       &       &       &       &       &      \\
lqscv &  0.58 &   1.00 &       &       &       &       &       &       &      \\
dbq   &  0.03 &   0.16 &  1.00 &       &       &       &       &       &      \\
dnf   &  0.13 &   0.06 & -0.02 &  1.00 &       &       &       &       &      \\
dno   &  0.03 &  -0.07 &  0.06 &  0.23 &  1.00 &       &       &       &      \\
snm   &  0.26 &  -0.05 & -0.09 &  0.22 &  0.17 &  1.00 &       &       &      \\
lss   &  0.14 &   0.16 &  0.00 &  0.44 &  0.43 &  0.48 &  1.00 &       &      \\
ddd   & -0.05 &   0.05 & -0.05 &  0.24 &  0.35 &  0.22 &  0.48 &  1.00 &      \\
dp    & -0.11 &   0.14 &  0.09 &  0.09 &  0.29 & -0.54 & -0.04 &  0.27 &  1.00 \\
\bottomrule
\end{tabular}
\end{table}

\renewcommand{\thetable}{\Alph{section}\arabic{table}}
\renewcommand{\thefigure}{\Alph{section}\arabic{figure}}
\setcounter{table}{0}
\setcounter{figure}{0}

\newpage

\section{Appendix Figures}

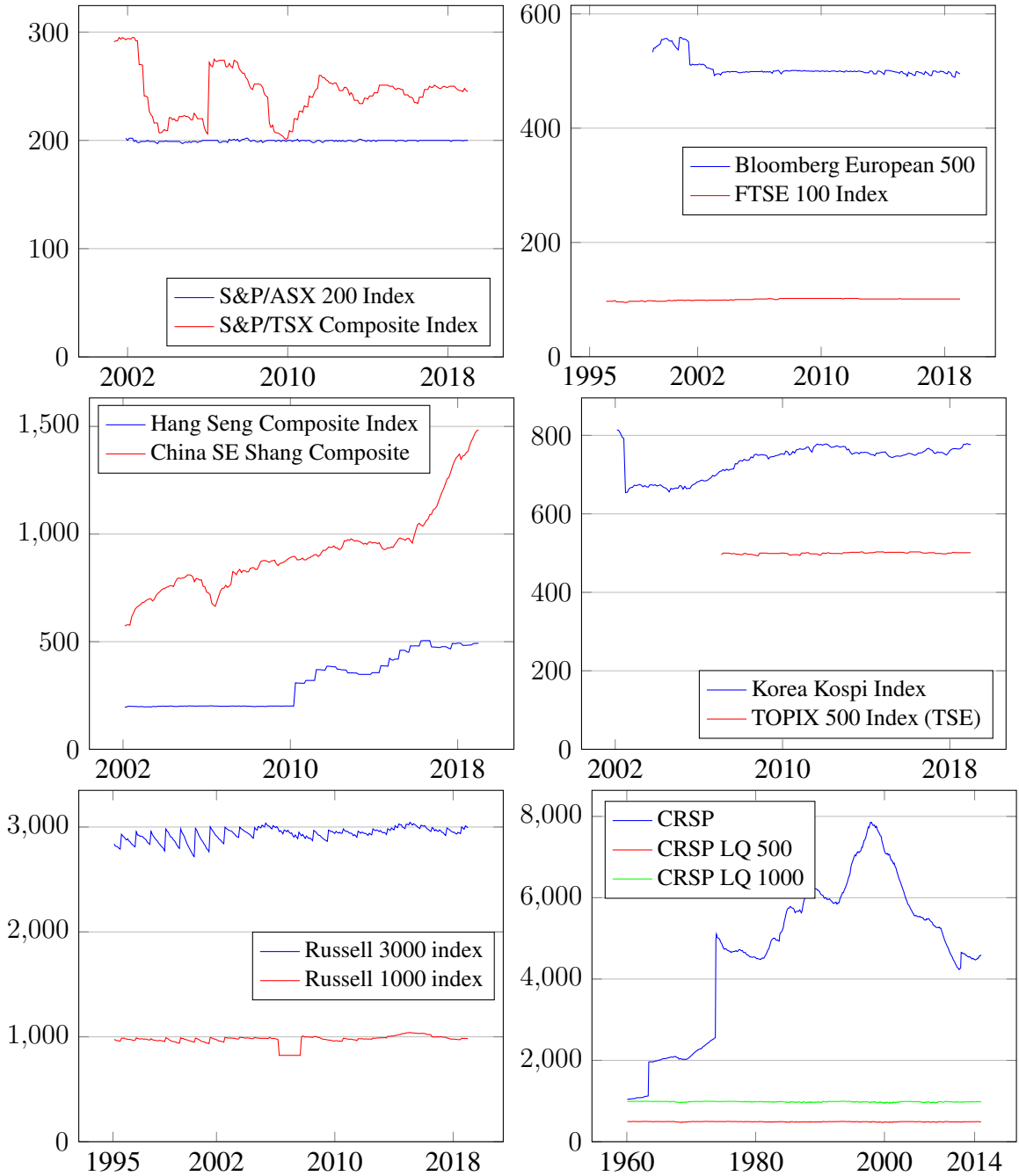
\begin{figure}[tbph]
    \centering
    \begin{tikzpicture}
        \begin{axis}[width=8cm, date coordinates in=x, ymin=0, xtick={1995-01-01, 2002-01-01, 2010-01-01, 2018-01-01}, xticklabel=\year, legend entries={S\&P/ASX 200 Index, S\&P/TSX Composite Index}, legend cell align={left}, legend pos=south east, legend style={font=\footnotesize}, ymajorgrids]
        \addplot [blue] table {data/size_as51_idx.dat};    
        \addplot [red] table {data/size_sptsx_idx.dat};    
        \end{axis}
    \end{tikzpicture}
    \begin{tikzpicture}
        \begin{axis}[width=8cm, date coordinates in=x, ymin=0, xtick={1995-01-01, 2002-01-01, 2010-01-01, 2018-01-01}, xticklabel=\year, legend entries={Bloomberg European 500, FTSE 100 Index}, legend cell align={left}, legend style={at={(0.98, 0.5)},anchor=east, font=\footnotesize}, ymajorgrids]
        \addplot [blue] table {data/size_be500_idx.dat};    
        \addplot [red] table {data/size_ukx_idx.dat};    
        \end{axis}
    \end{tikzpicture}
    \begin{tikzpicture}
        \begin{axis}[width=8cm, date coordinates in=x, ymin=0, xtick={1995-01-01, 2002-01-01, 2010-01-01, 2018-01-01}, xticklabel=\year, legend entries={Hang Seng Composite Index, China SE Shang Composite}, legend cell align={left}, legend style={at={(0.02, 0.98)},anchor=north west,font=\footnotesize}, ymajorgrids]
        \addplot [blue] table {data/size_hsci_idx.dat};    
        \addplot [red] table {data/size_shcomp_idx.dat};
        \end{axis}
    \end{tikzpicture}
    \begin{tikzpicture}
        \begin{axis}[width=8cm, date coordinates in=x, ymin=0, xtick={1995-01-01, 2002-01-01, 2010-01-01, 2018-01-01}, xticklabel=\year, legend entries={Korea Kospi Index, TOPIX 500 Index (TSE)}, legend cell align={left}, legend pos=south east, legend style={font=\footnotesize}, ymajorgrids]
        \addplot [blue] table {data/size_kospi_idx.dat};    
        \addplot [red] table {data/size_tpx500_idx.dat};    
        \end{axis}
    \end{tikzpicture}
    \begin{tikzpicture}
        \begin{axis}[width=8cm, date coordinates in=x, ymin=0, xtick={1995-01-01, 2002-01-01, 2010-01-01, 2018-01-01}, xticklabel=\year, legend entries={Russell 3000 index, Russell 1000 index}, legend cell align={left}, legend style={at={(0.98, 0.5)},anchor=east,font=\footnotesize}, ymajorgrids]
        \addplot [blue] table {data/size_ray_idx.dat};    
        \addplot [red] table {data/size_riy_idx.dat};    
    \end{axis}
    \end{tikzpicture}
    \begin{tikzpicture}
        \begin{axis}[width=8cm, date coordinates in=x, ymin=0, xtick={1960-01-01, 1980-01-01, 2000-01-01, 2014-01-01}, xticklabel=\year, legend entries={CRSP, CRSP LQ 500, CRSP LQ 1000}, legend cell align={left}, legend style={font=\footnotesize}, ymajorgrids, legend pos=north west]
        \addplot [blue] table {data/size_cfm_crspus_idx.dat};
        \addplot [red] table {data/size_cfm_crspus_lq500_t1.dat};
        \addplot [green] table {data/size_cfm_crspus_lq500_t12.dat};
    \end{axis}
    \end{tikzpicture}
    \caption{Number of constituents in each pool.}
    \label{fig:size}
\end{figure}

\newpage

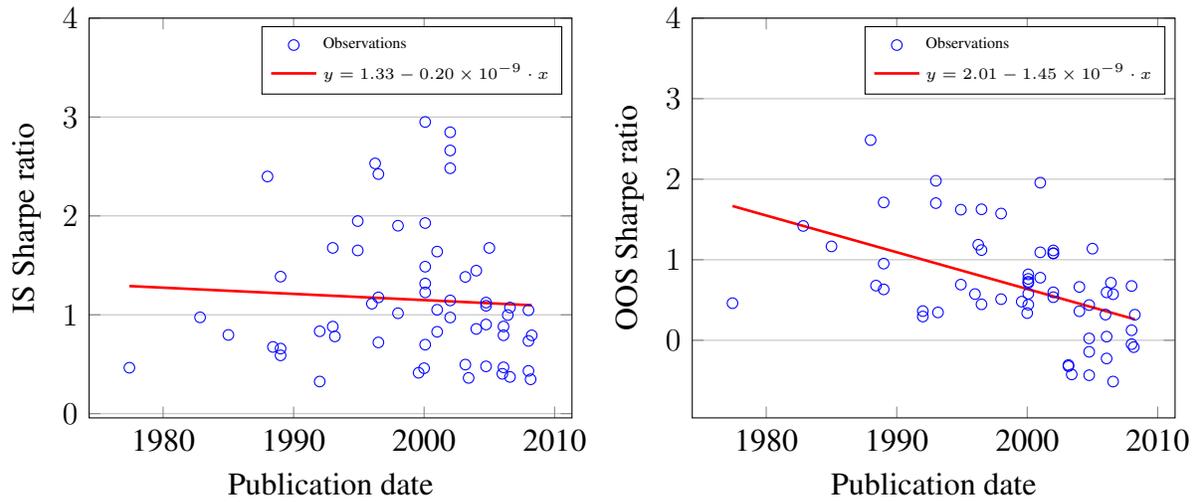
\begin{figure}[tbph]
    \centering
    \begin{tikzpicture}
        \begin{axis}[
            width=8cm,
            date coordinates in=x,
            xtick={1970-01-01, 1980-01-01, 1990-01-01, 2000-01-01, 2010-01-01}, 
            xticklabel=\year,
            ymajorgrids,
            ymax=4,
            xlabel={Publication date},
            ylabel={IS Sharpe ratio},
            legend entries={
              Observations,
              $y=1.33-0.20\times10^{-9}\cdot x$,
            },
            legend cell align={left},
            legend style={
              font=\tiny,
            },
        ]
            \addplot+ [blue, only marks, mark=o] table {data/is_dapublished_crsp.dat};            
            \addplot [red, line width=1pt] table {data/is_dapublished_crsp_reg.dat};
        \end{axis}
    \end{tikzpicture}
    \begin{tikzpicture}
        \begin{axis}[
            width=8cm,
            date coordinates in=x,
            xtick={1970-01-01, 1980-01-01, 1990-01-01, 2000-01-01, 2010-01-01}, 
            xticklabel=\year,
            ymajorgrids,
            ymax=4,
            xlabel={Publication date},
            ylabel={OOS Sharpe ratio},
            legend entries={
              Observations,
              $y=2.01-1.45\times10^{-9}\cdot x$,
            },
            legend cell align={left},
            legend style={
              font=\tiny,
            },
        ]
            \addplot+ [blue, only marks, mark=o] table {data/oos_dapublished_crsp.dat};            
            \addplot [red, line width=1pt] table {data/oos_dapublished_crsp_reg.dat};
        \end{axis}
    \end{tikzpicture}
    \caption{In-sample (resp. out-of-sample) Sharpe ratio as a function of publication date.
    Factors are beta-neutralised and computed on CRSP, conditional on its in-sample Sharpe ratio being greater than 0.3.
    One dot per factor.
    The red line draws a linear trend, fitted on blue dots, where dates are represented in Unix time.
    The $R^2$ are $0.4\%$ and $24\%$ respectively.}
    \label{fig:is_oos_dapublished}
\end{figure}

\end{document}